\documentclass{ws-mpla}

\usepackage{epsfig}
\usepackage{epsf}

\begin{document}

\markboth{L. SHAO and B.-Q. MA} {First Digit Distribution of Hadron
Full Width}

\catchline{}{}{}{}{}

\title{FIRST DIGIT DISTRIBUTION OF HADRON
FULL WIDTH\footnote{Published in Mod. Phys. Lett. A 24 (2009)
3275-3282}}

\author{\footnotesize LIJING SHAO}
\address{School of Physics and State Key Laboratory of Nuclear Physics and
Technology, \\
Peking University, Beijing 100871, China}

\author{\footnotesize BO-QIANG MA}

\address{School of Physics and State Key Laboratory of Nuclear Physics and
Technology, \\
Peking University, Beijing 100871, China\\
mabq@phy.pku.edu.cn}

\maketitle


\begin{abstract}
A phenomenological law, called Benford's law, states that the
occurrence of the first digit, i.e., $1,2,...,9$, of numbers from
many real world sources is not uniformly distributed, but instead
favors smaller ones according to a logarithmic distribution. We
investigate, for the first time, the first digit distribution of the
full widths of mesons and baryons in the well defined science domain
of particle physics systematically, and find that they agree
excellently with the Benford distribution. We also discuss several
general properties of Benford's law, i.e., the law is
scale-invariant, base-invariant, and power-invariant. This means
that the lifetimes of hadrons follow also Benford's law.

\end{abstract}

\ccode{PACS Nos.: 02.50.Cw; 14.20.-c; 14.40.-n; 21.10.Tg.}

\vspace{6mm}

Many numbers in the real world follow a fascinating pattern of
distribution rather than a uniform distribution as might be
expected. In 1881, Newcomb\cite{newcomb1881} noticed that the
preceding pages of the logarithmic table wear out faster, thus he
hinted at the idea that the first nonzero digit of many natural
numbers favors small values. Then in 1938, Benford\cite{benford1938}
investigated a great number of data sets in various unrelated
fields, e.g., the arabic numbers on the front page of a newspaper,
the wire and drill gauges of the mechanic, the magnitude scales of
astronomy, the street addresses published in a magazine, the weights
of molecules and atoms, the areas of lakes and the lengths of
rivers, and he found that they all agree with a logarithmic
distribution which now we refer to as Benford's law, or the first
digit law,
\begin{equation}\label{benford}
P(k) = \log_{10} (1 + \frac{1}{k}), \, k=1,2,...,9
\end{equation}
where $P(k)$ is the probability of a number having the first nonzero
digit $k$.

Surprisingly, there are plenty of unrelated data sets in the nature,
e.g., physical constants\cite{burke1991}, alpha-decay
half-lives\cite{buck1993,ni2008}, survival
distributions\cite{leemis2000}, the strengths of electric-dipolar
lines in transition arrays of complex atomic spectra\cite{pain2008},
and numbers in our everyday lives, e.g., the stock market
indices\cite{ley1996}, numbers in the World Wide
Web\cite{dorogovtsev2006}, the file sizes in a personal
computer\cite{torres2007}, the winning bids for certain eBay
auctions\cite{giles2007}, all incredibly conform to the peculiar
first digit law perfectly. Furthermore, Tolle et al.\cite{tolle2000}
provided empirical evidence that dynamical systems reveal the first
digital behavior, and then Snyder et al.\cite{snyder2001} developed
corresponding computer simulations and confirmed the conclusion, and
later it was discussed further by Berger et al.\cite{berger2005} in
detail. Nevertheless, there also exist many data sets which do not
obey the law, and unfortunately there is no priori criteria yet to
judge which type a data set belongs to.

After the discovery for more than one century since 1881, a
universally accepted explanation for the underlying reason of such a
law is still lacking. However, many literatures have exploited a lot
of applications of the logarithmic distribution in various fields,
and the first digit law has been used in practice already. The main
usage is to detect data and judge their reasonableness. It is
applied in distinguishing and ascertaining fraud in taxing and
accounting\cite{nigrini1996,nigrini1999,rose2003}, fabrication in
clinical trials\cite{marzouki2005}, authenticity of the pollutant
concentrations in ambient air\cite{brown2005}, electoral cheats or
voting anomalies\cite{torres2007}, and falsified data in scientific
experiments\cite{diekmann2007}.
Moreover, the first digit law is largely applied in computer science
for speeding up calculation\cite{barlow1985}, minimizing expected
storage space\cite{schatte1988}, analyzing the behavior of
floating-point arithmetic algorithms\cite{berger2007}, and
especially for various studies in the image
domain\cite{jolion2001,fu2007}.

Despite so many successful developments, there is no academic paper
yet on the topic of Benford's law in the well defined science domain
of particle physics, to the best of our knowledge. For the first
time, we investigate the properties of particles by applying the
first digit law, and find that in spite of the smallness of data
capacity and the large span of the ranges, e.g., $\Gamma \sim
10^{-24}$~MeV for the neutron and $\Gamma \sim 10^3$~MeV for ${\rm
f}_0 (600)$), the full widths of mesons and baryons pertain to the
Benford distribution very well respectively. Hence the Benford's law
applies to the full widths of hadrons with a rather good precision.

\begin{figure}
\begin{center}
\scalebox{1.2}{\includegraphics{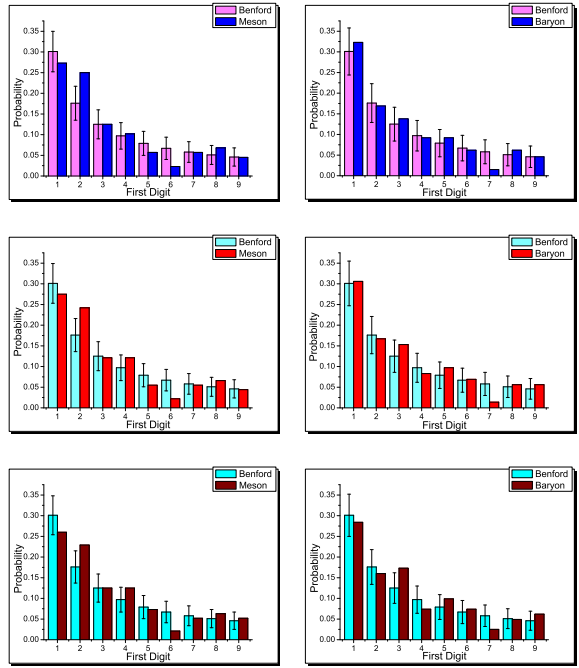}}\caption{Comparisons of
Benford's law and the distribution of the first digit of the full
widths of mesons (left) and baryons (right).}\label{layout}
\end{center}
\end{figure}

More specifically, we scrutinize the meson summary table and baryon
summary table in {Review of Particle Physics} (2008) by Particle
Data Group\cite{PDG}, and compare the first digit distribution of
the full widths of mesons and baryons with Benford distribution
respectively. The systematical statistical results are shown in
Table~\ref{meson} for mesons and Table~\ref{baryon} for baryons,
respectively. The numbers in the bracket are the expected number,
\begin{equation}\label{nben}
N_{\rm Ben} = N \, \log_{10}(1+1/k)
\end{equation}
together with the root mean square error evaluated by the binomial
distribution,
\begin{equation}\label{deltan}
\Delta N = \sqrt{N \, P(k) \,(1-P(k))}.
\end{equation}
The detail of the classification from Case 1 to Case 3 will be
discussed later. We can see that the results are very weakly case
sensitive, and all are in good agreement with Benford's law. The
intuitive figures are illustrated in Fig.~\ref{layout}, with the
left row for mesons, the right row for baryons for Case 1 to Case 3
from the top down.

\begin{table}
\tbl{The first digit distribution of the full widths of mesons.}
{\begin{tabular}{c||cl|cl|cl}
\hline %
{\bf First Digit} & \multicolumn{2}{c|}{\bf Case 1 (88)}&
\multicolumn{2}{c|}{\bf Case 2 (91)} &\multicolumn{2}{c}{\bf Case 3 (96)}\\
\hline %

1 & 24 &(26.5$\pm$4.3) & 25 &(27.4$\pm$4.4) & 25 &(28.9$\pm$4.5)\\
2 & 22 &(15.5$\pm$3.6) & 22 &(16.0$\pm$3.6) & 22 &(16.9$\pm$3.7)\\
3 & 11 &(11.0$\pm$3.1) & 11 &(11.4$\pm$3.2) & 12 &(12.0$\pm$3.2)\\
4 & 9  & (8.5$\pm$2.8) & 11 &(8.8$\pm$2.8)  & 12 &(9.3$\pm$2.9) \\
5 & 5  & (7.0$\pm$2.5) & 5  &(7.2$\pm$2.6)  & 7  &(7.6$\pm$2.6) \\
6 & 2  & (5.9$\pm$2.3) & 2  &(6.1$\pm$2.4)  & 2  &(6.4$\pm$2.4) \\
7 & 5  & (5.1$\pm$2.2) & 5  &(5.3$\pm$2.2)  & 5  &(5.6$\pm$2.3) \\
8 & 6  & (4.5$\pm$2.1) & 6  &(4.7$\pm$2.1)  & 6  &(4.9$\pm$2.2) \\
9 & 4  & (4.0$\pm$2.0) & 4  &(4.2$\pm$2.0)  & 5  &(4.4$\pm$2.0) \\

{\bf Pearson} $\mathbf{\chi^2}$ & \multicolumn{2}{c|}{\bf 6.62}
& \multicolumn{2}{c|}{\bf 6.82} & \multicolumn{2}{c}{\bf 6.32}\\
\hline %
\end{tabular}\label{meson}}
\end{table}

\begin{table}
\tbl{The first digit distribution of the full widths of baryons.}
{\begin{tabular}{c||cl|cl|cl}
\hline %
{\bf First Digit} & \multicolumn{2}{c|}{\bf Case 1 (65)} &
\multicolumn{2}{c|}{\bf Case 2 (72)} &\multicolumn{2}{c}{\bf Case 3 (81)}\\
\hline %

1 & 21 &(19.6$\pm$3.7) & 22 &(21.7$\pm$3.9) & 23 &(24.4$\pm$4.1)\\
2 & 11 &(11.4$\pm$3.1) & 12 &(12.7$\pm$3.2) & 13 &(14.3$\pm$3.4)\\
3 & 9  &(8.1$\pm$2.7)  & 11 &(9.0$\pm$2.8)  & 14 &(10.1$\pm$3.0)\\
4 & 6  & (6.3$\pm$2.4) & 6  &(7.0$\pm$2.5)  & 6  &(7.8$\pm$2.7) \\
5 & 6  & (5.1$\pm$2.2) & 7  &(5.7$\pm$2.3)  & 8  &(6.4$\pm$2.4) \\
6 & 4  & (4.4$\pm$2.0) & 5  &(4.8$\pm$2.1)  & 6  &(5.4$\pm$2.2) \\
7 & 1  & (3.8$\pm$1.9) & 1  &(4.2$\pm$2.0)  & 2  &(4.7$\pm$2.1) \\
8 & 4  & (3.3$\pm$1.8) & 4  &(3.7$\pm$1.9)  & 4  &(4.1$\pm$2.0) \\
9 & 3  & (3.0$\pm$1.7) & 4  &(3.3$\pm$1.8)  & 5  &(3.7$\pm$1.9) \\

{\bf Pearson} $\mathbf{\chi^2}$ & \multicolumn{2}{c|}{\bf 2.57}
& \multicolumn{2}{c|}{\bf 3.52} & \multicolumn{2}{c}{\bf 4.57}\\
\hline %
\end{tabular}\label{baryon}}
\end{table}

It is worthy to mention that, when estimating the fitness to the
theoretical probability distribution, we should use fitness
estimating $\chi^2$, namely Pearson $\chi^2$,
\begin{equation}\label{pearson}
\chi^2(n-1) = \sum_{i=1}^{n} \frac{(N_{\rm Obs} - N_{\rm
Ben})^2}{N_{\rm Ben}}
\end{equation}
where $N_{\rm Obs}$ is the observational number and $N_{\rm Ben}$ is
the theoretical number from Benford's law, and here in our question
$n=9$. However, it is not appropriate to use parameter estimating
$\chi^2$ as used in Refs.~\refcite{buck1993,ni2008}. In
Eq.~(\ref{pearson}), the degree of freedom is $9-1=8$, and under the
confidence level 95\%, $\chi^2(8) = 15.507$, and under the
confidence level 50\%, $\chi^2(8) = 7.344$. The $\chi^2$ we
calculated is smaller than those in Refs.~\refcite{buck1993,ni2008},
indicating clearly that the fitness is remarkably good in particle
physics.

The classification from Case 1 to Case 3 is due to the
incompleteness and uncertainty of experimental data, however, we do
our best to treat data without bias. Since we only deal with the
full widths of hadrons, we ignore the $e^+ e^-$ width $\Gamma_{\rm
ee}$ and we do not distinguish the difference between the full width
and the Breit-Wigner full width given in the baryon summary table.
When the summary tables give only the lifetime $\tau$ instead of the
full width $\Gamma$, we use $\Gamma\times\tau=\hbar$ to get the
corresponding full width. And we drop the single-side data, e.g.,
$\Gamma < 1.9$~MeV for $\Lambda_c(2625)^+$, while we still keep the
double-side data and pick the mean value of the boundaries, for
instance, given 200 to 500~MeV for ${\rm f}_0(1370)$, we treat it as
$\Gamma=350$~MeV, and when given the most likely values in these
two-boundary cases, we chose them for simplicity. But there still
remain some puzzled data due to the isospin problem. For most
hadrons of the same isospin $I$, there are several types of
particles due to the different isospin projection $I_3$, and the
summary table separates some while it does not distinguish between
others definitely, however, the same isospin does not always promise
the same lifetime hence the same full width. For the stringency of
our approach, we use the following three schemes to deal with the
data:
\begin{itemlist}
 \item In Case 1, we just drop out the puzzled data;
 \item In Case 2, we
faithfully follow the classification published in the particle table
and treat each item as a whole, and when there are several full
widths appearing under one item, we pick the mean value and only
count for once;
 \item And in Case 3, we stick to the appearance of the
data, that is, when there is a datum, we count it for once.
\end{itemlist}

\begin{table}
\tbl{The first digit distribution of the full widths of hadrons.}
{\begin{tabular}{c||cl|cl|cl}
\hline %
{\bf First Digit} & \multicolumn{2}{c|}{\bf Case 1 (153)}&
\multicolumn{2}{c|}{\bf Case 2 (163)} &\multicolumn{2}{c}{\bf Case 3 (177)}\\
\hline %

1 & 45 &(46.1$\pm$5.7) & 47 &(49.1$\pm$5.9) & 48 &(53.3$\pm$6.1)\\
2 & 33 &(26.9$\pm$4.7) & 34 &(28.7$\pm$4.9) & 35 &(31.2$\pm$5.1)\\
3 & 20 &(19.1$\pm$4.1) & 22 &(20.4$\pm$4.2) & 26 &(22.1$\pm$4.4)\\
4 & 15 &(14.8$\pm$3.7) & 17 &(15.8$\pm$3.8) & 18 &(17.2$\pm$3.9)\\
5 & 11 &(12.1$\pm$3.3) & 12 &(12.9$\pm$3.4) & 15 &(14.0$\pm$3.6)\\
6 & 6  &(10.2$\pm$3.1) & 7  &(10.9$\pm$3.2) & 8  &(11.8$\pm$3.3)\\
7 & 6  &(8.9$\pm$2.9)  & 6  &(9.5$\pm$3.0)  & 7  &(10.3$\pm$3.1)\\
8 & 10 &(7.8$\pm$2.7)  & 10 &(8.3$\pm$2.8)  & 10 &(9.1$\pm$2.9) \\
9 & 7  &(7.0$\pm$2.6)  & 8  &(7.5$\pm$2.7)  & 10 &(8.1$\pm$2.8) \\

{\bf Pearson} $\mathbf{\chi^2}$ & \multicolumn{2}{c|}{\bf 4.82}
& \multicolumn{2}{c|}{\bf 4.39} & \multicolumn{2}{c}{\bf 4.62}\\
\hline %
\end{tabular}\label{hadron}}
\end{table}

\begin{figure}
\begin{center}
\scalebox{0.5}{\includegraphics{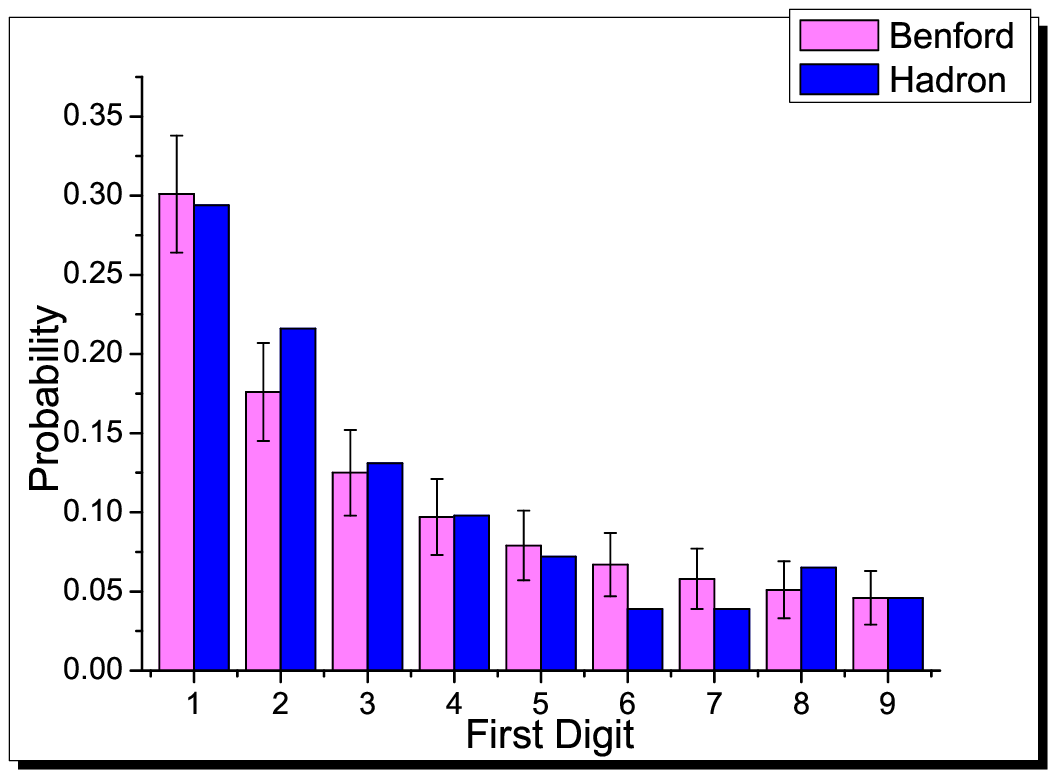}}\caption{
Comparison of Benford's law and the first digit distribution of the
full widths of hadrons in Case 1.}\label{ghadronwithout}
\scalebox{0.5}{\includegraphics{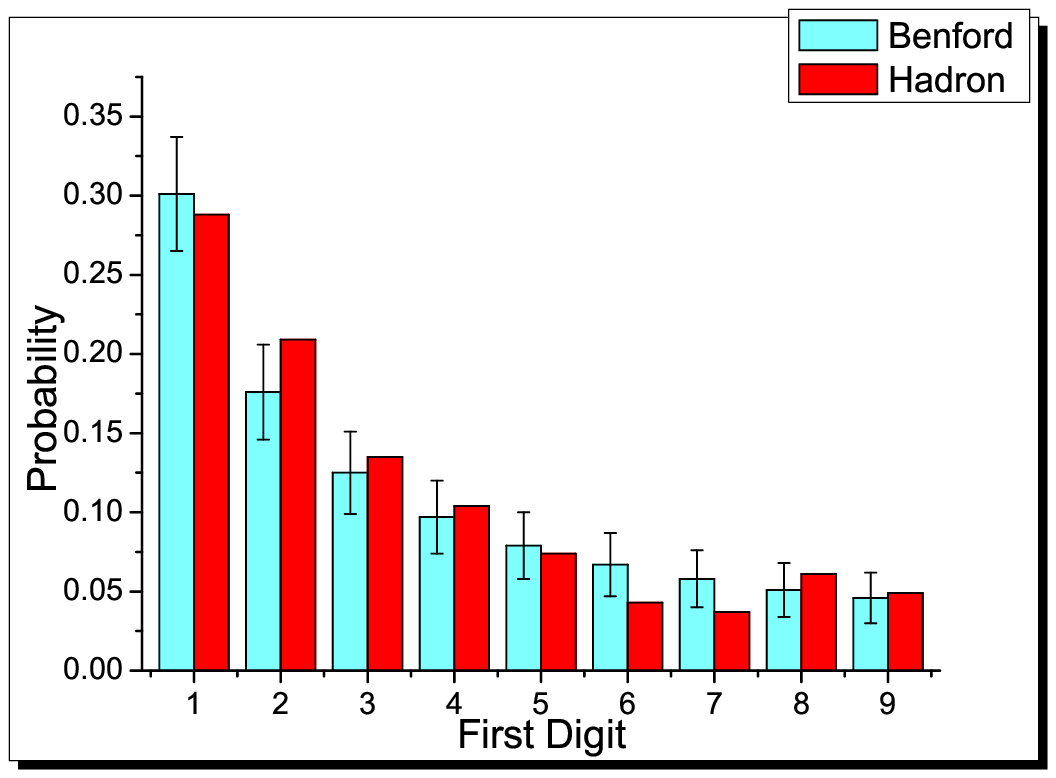}}\caption{
Comparison of Benford's law and the first digit distribution of the
full widths of hadrons in Case 2.}\label{ghadronaverage}
\scalebox{0.5}{\includegraphics{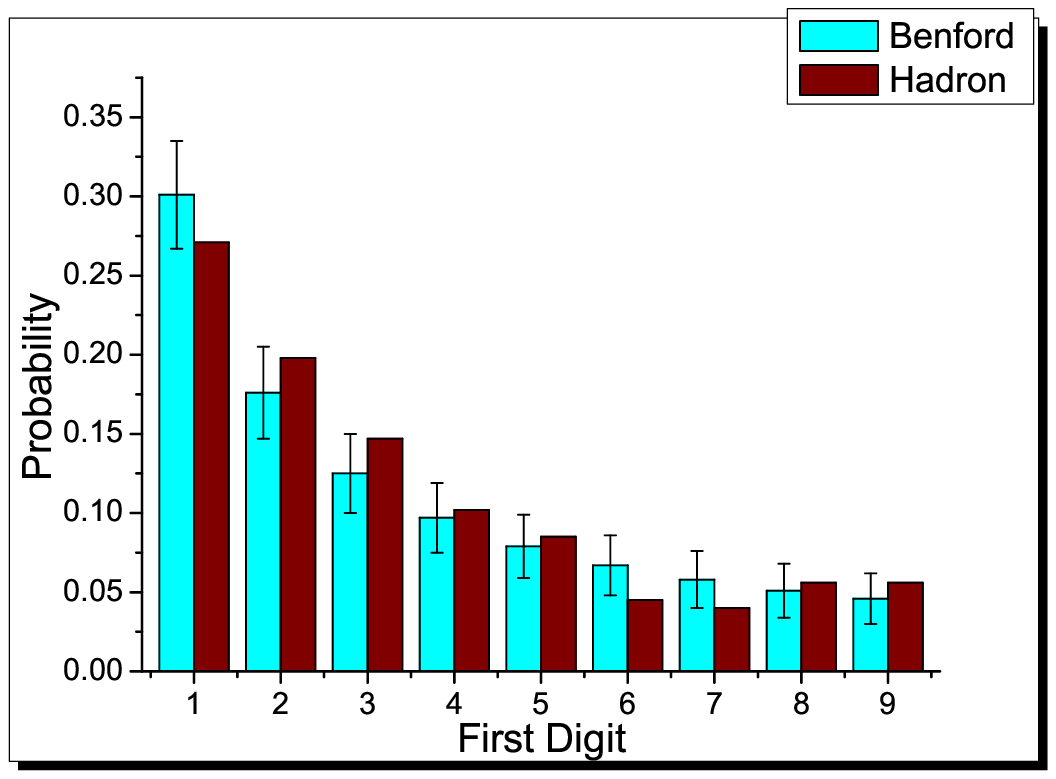}}\caption{
Comparison of Benford's law and the first digit distribution of the
full widths of hadrons in Case 3.}\label{ghadronwithall}
\end{center}
\end{figure}

Though the fitness is so impressive, we still feel short of data.
Consequently, we add mesons and baryons up to get the distribution
of hadrons. The results are shown in Table~\ref{hadron} numerically,
and in Fig.~\ref{ghadronwithout}, Fig.~\ref{ghadronaverage} and
Fig.~\ref{ghadronwithall} graphically. They all appear remarkably
good.

Many attempts have been tried to explain the underlying reason for
Benford's law. For theoretical reviews, see Ref.~\refcite{raimi1976}
and papers written by
Hill\cite{hill1995at102,hill1995at123,hill1995at10}. We here present
some discussions on the first digit law, focusing on its several
general properties.

Firstly, a positive date set can be rewritten in the form of \{$10^{
n_i + f_i }$\}, where $n_i$ is an integer not affecting the first
digit, and $f_i$ ($0 \leq f_i < 1$) is the fractional part which
contributes to the question. As suggested and derived by
Newcomb\cite{newcomb1881}, a uniform distribution of the fractional
part $f_i$ of the exponent in the interval $[0,1)$ leads to the
logarithmic law.

Secondly, as can be expected, multiplication of a constant upon the
data set does not change the probability distribution, the reason of
which is that the exponent of new datum $N^\prime$ satisfies
\begin{equation}\label{scale}
\log_{10} N^\prime = \log_{10} (C \times N) = \log_{10}C + \log_{10}
N.
\end{equation}
This property means that the law does not depend on any particular
choice of units, namely scale-invariance\cite{pinkham1961}, which
was discovered by Pinkham in 1961. In mathematics, this law is the
only digital law that is scale-invariant.

Thirdly, any power of the data set does not change the distribution
either, i.e., the law is power-invariant. Because of $\log_{10}
N^\prime = \log_{10} N^\alpha = \alpha \log_{10} N$ for $\alpha \neq
0$, the logarithmic distribution remains unchanged according to the
first property. In our research, the full widths of hadrons fit the
logarithmic distribution, so do the lifetimes of hadrons.

Fourthly, the Benford's law is
base-invariant\cite{hill1995at102,hill1995at123,hill1995at10} too,
which means that it is independent of the base $d$ you use. In the
binary system ($d$=2), octal system ($d$=8), or other base system,
the data, as well as in the decimal system ($d$=10), all fit the
general first digit law,
\begin{equation}
P(k) = \log_{d}(1+\frac{1}{k}), \, k=1, 2, ..., {d-1}.
\end{equation}
Hill proved strictly that ``scale-invariance implies
base-invariance''\cite{hill1995at102} and ``base-invariance implies
Benford's law''\cite{hill1995at123} mathematically in the framework
of probability theory.

Fifthly, in 2001, Pietronero et al.\cite{pietronero2001} provided a
new insight, suggesting that a process or an object $N(t)$ with its
time evolution governed by multiplicative fluctuations generates
Benford's law naturally and they used stockmarket as a convictive
example. Further, they demonstrated it with the computer simulation
and got rather good result. The main idea is that $N(t + \delta t) =
r(t) \times N(t)$, where $r(t)$ is a random variable. After treating
$\log r(t)$ as a new random variable, it is a Brownian process $\log
N(t + \delta t) = \log r(t) + \log N(t)$ in the logarithmic space.
Utilizing the central limit theorem in a large sample, $\log N(t)$
becomes uniformly distributed. Thus,
\begin{equation}\label{multi}
P(k) = \frac{\int_{k}^{k+1} {\rm d} \log N(t)}{\int_{1}^{10} {\rm d}
\log N(t)} = \log_{10} (1 + \frac{1}{k}),
\end{equation}
which is exactly the formula of Benford's law given in
Eq.~(\ref{benford}). This approach is well recommended in
Refs.~\refcite{ni2008,li2004}.


In summary, we applied Benford's law to the well defined science
domain of particle physics for the first time. The distribution of
the first digits of the full widths of hadrons, including mesons and
baryons respectively, all fit the logarithmic law remarkably well.
Moreover, we discussed several general properties of the law, and
reached the conclusion that our results apply to the lifetimes of
hadrons as well. It is still a challenge to find the basic reason
for the common distribution pattern among various sorts of natural
behaviors. Our results suggest the necessity to look into some
hitherto unnoticed features of basic physical phenomena. The first
digit law can serve as a tool to test the reasonableness of any
theory or model that is supposed to be the underlaying theory of the
nature.

\section*{Acknowledgments}
This work is partially supported by National Natural Science
Foundation of China (Nos.~10721063, 10575003, 10528510), by the Key
Grant Project of Chinese Ministry of Education (No.~305001). It is
also supported by Hui-Chun Chin and Tsung-Dao Lee Chinese
Undergraduate Research Endowment (Chun-Tsung Endowment) at Peking
University, and by National Fund for Fostering Talents of Basic
Science (No.~J0630311 and No.~J0730316).



\end{document}